\documentclass[11pt]{article}

\usepackage{amsmath,amssymb}
\usepackage{amsthm}
\usepackage{ascmac}
\usepackage{bm}
\usepackage{float}

\usepackage{graphicx}
\usepackage[all]{xy}

\usepackage[top=25truemm,bottom=25truemm,left=30truemm,right=30truemm]{geometry}

\numberwithin{equation}{section}

\begin{document}

\renewcommand{\include}[1]{}
\renewcommand\documentclass[2][]{}

\renewcommand{\thefootnote}{\fnsymbol{footnote}}
\setcounter{page}{0}
\thispagestyle{empty}
\begin{flushright} OU-HET 899 \end{flushright} 

\vskip3cm
\begin{center}
{\LARGE {\bf Geometric transition in Non-perturbative Topological string}}  
\vskip1.5cm
{\large 
{\bf Yuji Sugimoto\footnote{sugimoto@het.phys.sci.osaka-u.ac.jp}} 

\vskip1cm
\it Department of Physics, Graduate School of Science, Osaka University, \\ Toyonaka, Osaka 560-0043, Japan} 
\end{center}

\vskip1cm
\begin{abstract} 
We study a geometric transition in non-perturbative topological string. We consider two cases.  One is the geometric transition from the closed topological string on the local $\mathcal{B}_{3}$ to the closed topological string on the resolved conifold. The other is the geometric transition from the closed topological string on the local $\mathcal{B}_{3}$ to the open topological string on the resolved conifold with a toric A-brane. We find that, in both cases, the geometric transition can be applied for the non-perturbative topological string. We also find the corrections of the value of K\"ahler parameters at which the geometric transition occurs.
\end{abstract}

\renewcommand{\thefootnote}{\arabic{footnote}}
\setcounter{footnote}{0}

\vfill\eject

\hrulefill
\tableofcontents
\hrulefill

\section{Introduction}
Recently, in unrefined topological string theory  on non-compact toric Calabi-Yau threefolds, the free energy including non-perturbative effects is proposed \cite{Hatsuda:2013oxa}. The non-perturbative parts can be obtained by considering the Nekrasov--Shatashvili limit  \cite{Nekrasov:2009rc}. We call this as ``non-perturbative free energy". The non-perturbative free energy is finite for any string coupling by the HMO cancellation mechanism \cite{Hatsuda:2012dt}. This free energy has been studied in various situations \cite{Hatsuda:2013yua}\cite{Hirano:2014bia}\cite{Kallen:2013qla}\cite{Huang:2014eha}\cite{Moriyama:2014nca}\cite{Hatsuda:2015oaa}\cite{Hatsuda:2016uqa}\cite{Couso-Santamaria:2014iia}\cite{Krefl:2015vna}\cite{Wang:2014ega}\cite{Wang:2015wdy}. Especially, it is known that this free energy is closely related to the quantization of a mirror curve for the non-compact toric Calabi-Yau threefold \cite{Marino:2015ixa}\cite{Kashaev:2015wia}\cite{Grassi:2014zfa}\cite{Hatsuda:2015fxa}\cite{Codesido:2015dia}\cite{Kashani-Poor:2016edc}\cite{Hatsuda:2016rmv}. This provides us with the non-perturbative definition of the topological string. However, it is unclear  whether the important properties of the perturbative topological string hold, even if we consider the non-perturbative topological string.
\par
In this paper, we study a geometric transition \cite{Gopakumar:1998ii}\cite{Gopakumar:1998jq}\cite{Gopakumar:1998ki} in the non-perturbative topological string. As an example, we study the geometric transition in the closed topological string on the local $\mathcal{B}_{3}$. We consider two cases. One is the geometric transition from the local $\mathcal{B}_{3}$ to the resolved conifold in the closed topological string. The other is the geometric transition from the closed topological string on the local $\mathcal{B}_{3}$ to the open topological string on the resolved conifold with a toric A-brane.
\par
We first consider the geometric transition from the local $\mathcal{B}_{3}$ to the resolved conifold in the closed topological string. Then, we find that, by calculating the non-perturbative free energy of the closed topological string on the local $\mathcal{B}_{3}$ and the resolved conifold, the geometric transition can be applied even if the non-perturbative effects are included. We also find that, in contrast with the case of the perturbative free energy, the values of the K\"ahler parameters which the geometric transition occurs are corrected by the non-perturbative effects. 
\par
Next, we consider the geometric transition from the closed topological string to the open topological string with a toric A-brane. Then, we find that the HMO cancellation mechanism can be applied, even if there is a toric A-brane. We also find that the non-perturbative parts of this free energy have the same structure as the one in the references \cite{Kashani-Poor:2016edc}\cite{Aganagic:2011sg}\cite{Aganagic:2012hs}\cite{Marino:2016rsq}. We check this statement up to $\mathcal{O}(Q^2_{b})$. The K\"ahler parameters are corrected by the non-perturbative effects as in the above case.
\par
This paper is organized as follows. In section 2, we calculate the non-perturbative free energy of the closed topological string on the local $\mathcal{B}_{3}$ by using the refined topological vertex formalism \cite{Aganagic:2002qg}\cite{Aganagic:2003db}\cite{Iqbal:2007ii}\cite{Taki:2007dh}\cite{Awata:2009yc}\cite{Awata:2011ce}\cite{Bao:2013pwa}\cite{Hayashi:2013qwa}. In section 3, we consider the geometric transition from the local $\mathcal{B}_{3}$ to the resolved conifold in the closed topological string. We also consider the geometric transition which the toric A-brane occurs.  Finally, we summarize  our result and discuss the future work in section 4.

\section{Free energy for topological string on Local $\mathcal{B}_{3}$}
In this section, we calculate the non-perturbative free energy of the closed topological string on the local $\mathcal{B}_{3}$. We also check the HMO cancellation mechanism for this free energy.

\subsection{Calculation of Refined topological string}
In order to calculate the non-perturbative free energy, we use the refined topological vertex formalism. The web diagram of the local $\mathcal{B}_{3}$ is shown in Fig.\ref{localB3}, where we define $Q_{1,2,b,f}:=\mathrm{e}^{-t_{1,2,b,f}}$, and $t_{1,2,b,f}$ are the K\"ahler parameters.
\par
\begin{figure}[htbp]
\centering
    \includegraphics[width=15cm]{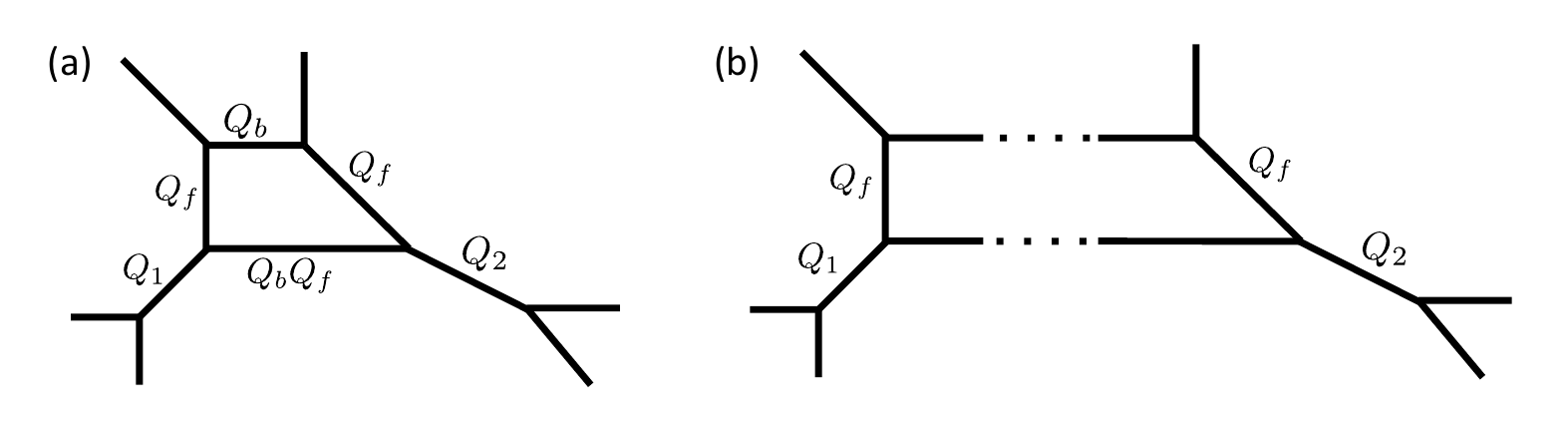}
    \caption{(a) is the web diagram of the local $\mathcal{B}_{3}$. (b) is its building blocks.}
    \label{localB3}
\end{figure}
Then, we can write the partition function of the refined topological string on this geometry $\mathcal{Z}_{\text{Local}~\mathcal{B}_{3}}(Q;t,q)$  as follow:
\begin{eqnarray}
\mathcal{Z}_{\text{Local}~\mathcal{B}_{3}}(Q;t,q)
=
\sum_{\mu_{b},\tilde{\mu}_{b}}
(-Q_{b})^{|\mu_{b}|}(-Q_{b}Q_{f})^{|\tilde{\mu}_{b}|}
f^2_{\tilde{\mu}_{b}}(t,q)
\mathcal{Z}^{(1)}_{\mu_{b}\tilde{\mu}_{b}}\mathcal{Z}^{(2)}_{\mu_{b}\tilde{\mu}_{b}},
\end{eqnarray}
where we define
\begin{eqnarray}
\mathcal{Z}^{(1)}_{\mu_{b}\tilde{\mu}_{b}}
&:=&
\sum_{\mu_{f},\mu_{1}}
(-Q_{f})^{|\mu_{f}|}(-Q_{1})^{|\mu_{1}|}
\tilde{f}_{\mu_{f}}(t,q)
C_{\emptyset \mu_{f} \mu_{b}}(t,q)C_{\mu^{t}_{f} \mu_{1} \tilde{\mu}_{b}}(t,q)
C_{\emptyset \mu^{t}_{1} \emptyset}(q,t),
\label{z1}
\\
\mathcal{Z}^{(2)}_{\mu_{b}\tilde{\mu}_{b}}
&:=&
\sum_{\mu_{f},\mu_{2}}
(-Q_{f})^{|\mu_{f}|}(-Q_{2})^{|\mu_{2}|}
\tilde{f}^{-1}_{\mu_{f}}(t,q)
C_{\mu_{f} \emptyset \mu^{t}_{b}}(q,t)C_{\mu_{2} \mu_{f}^{t} \tilde{\mu}^{t}_{b}}(q,t)
C_{\mu^{t}_{2} \emptyset \emptyset}(t,q).
\label{z2}
\end{eqnarray}
(\ref{z1}) and (\ref{z2}) correspond to the building blocks of the left side and the right side in Fig.\ref{localB3}(b), respectively. We also define the refined topological vertex $C_{\lambda \mu \nu}(t,q)$ and the framing factors $f_{\mu} (t,q)$, $\tilde{f}_{\mu} (t,q)$ as follows:
\begin{eqnarray}
C_{\lambda \mu \nu}(t,q) &=& t^{-\frac{||\mu^{t}||^{2}}{2}}q^{\frac{||\mu||^2 + ||\nu||^{2}}{2}} \tilde{Z}_{\nu}(t,q)\sum_{\eta}\Bigl(\frac{q}{t} \Bigr)^{\frac{|\eta| + |\lambda| - |\mu|}{2}}s_{\lambda^{t}/\eta}(t^{-\rho}q^{-\nu})s_{\mu/\eta}(t^{-\nu^{t}}q^{-\rho}),~~~~
\\ 
\tilde{Z}_{\nu}(t,q) &=& \prod_{(i,j) \in \nu}(1-q^{\nu_{i}-j}t^{\nu_{j}^{t} -i +1})^{-1},\\
f_{\mu} (t,q) &=& (-1)^{|\mu|}
q^{-\frac{||\mu||^2}{2}}t^{\frac{||\mu^{t}||^2}{2}},\\
\tilde{f}_{\mu} (t,q) &=& (-1)^{|\mu|}
\Bigl(\frac{t}{q}\Bigr)^{\frac{|\mu|}{2}}
q^{-\frac{||\mu||^2}{2}}t^{\frac{||\mu^{t}||^2}{2}}
,
\end{eqnarray}
where the function $s_{\lambda/\eta}(x_{1},x_{2},...)$ is the skew Schur function. By using some formulae in appendix, we obtain
\begin{eqnarray}
\mathcal{Z}^{(1)}_{\mu_{b}\tilde{\mu}_{b}}
&=&
\tilde{Z}_{\mu_{b}}(t,q)\tilde{Z}_{\tilde{\mu}_{b}}(t,q)q^{\frac{||\mu_{b}||^2 + ||\tilde{\mu}_{b}||^2}{2}}
\nonumber \\
&&\times
\prod_{i,j=1}^{\infty}
\frac{
(1-Q_{1}t^{-\tilde{\mu}^{t}_{b,j}+i-\frac{1}{2}}q^{j-\frac{1}{2}})
(1-Q_{1}Q_{f}t^{-\mu^{t}_{b,j}+i-\frac{1}{2}}q^{j-\frac{1}{2}})
}{
(1-Q_{f}t^{-\mu^{t}_{b,j}+i}q^{-\tilde{\mu}_{b,i}+j-1})
},
\label{unnorm1}
~~\\
\mathcal{Z}^{(2)}_{\mu_{b}\tilde{\mu}_{b}}
&=&
\tilde{Z}_{\mu^{t}_{b}}(q,t)\tilde{Z}_{\tilde{\mu}^{t}_{b}}(q,t)t^{\frac{||\mu^{t}_{b}||^2 + ||\tilde{\mu}^{t}_{b}||^2}{2}}
\nonumber \\
&&\times
\prod_{i,j=1}^{\infty}
\frac{
(1-Q_{2}t^{-\tilde{\mu}^{t}_{b,j}+i-\frac{1}{2}}q^{j-\frac{1}{2}})
(1-Q_{2}Q_{f}t^{-\mu^{t}_{b,j}+i-\frac{1}{2}}q^{j-\frac{1}{2}})
}{
(1-Q_{f}t^{-\mu^{t}_{b,j}+i-1}q^{-\tilde{\mu}_{b,i}+j})
}.
\label{unnorm2}
~~
\end{eqnarray}
In order to clarify the discussion in next section, we normalize (\ref{unnorm1}) and (\ref{unnorm2}) by dividing the trivial building blocks $\mathcal{Z}^{(1)}_{\emptyset \emptyset}$ and $\mathcal{Z}^{(2)}_{\emptyset \emptyset}$. Again by using some formulae, we obtain
\par
\begin{eqnarray}
\hat{\mathcal{Z}}^{(1)}_{\mu_{b}\tilde{\mu}_{b}}
&:=& \mathcal{Z}^{(1)}_{\mu_{b}\tilde{\mu}_{b}}/\mathcal{Z}^{(1)}_{\emptyset \emptyset}
\nonumber \\
&=&
\tilde{Z}_{\mu_{b}}(t,q)\tilde{Z}_{\tilde{\mu}_{b}}(t,q)q^{\frac{||\mu_{b}||^2 + ||\tilde{\mu}_{b}||^2}{2}}
\nonumber \\
&&\times
\prod_{(i,j)\in \mu_{b}}
\frac{
1-Q_{1}Q_{f}t^{-i+\frac{1}{2}}q^{\mu_{b,i}-j+\frac{1}{2}}
}{
1-Q_{f}t^{\tilde{\mu}^{t}_{b,j}-i+1}q^{\mu_{b,i}-j}
}
\prod_{(i,j)\in \tilde{\mu}_{b}}
\frac{
1-Q_{1}t^{-i+\frac{1}{2}}q^{\tilde{\mu}_{b,i}-j+\frac{1}{2}}
}{
1-Q_{f}t^{-\mu^{t}_{b,j}+i}q^{-\tilde{\mu}_{b,i}+j-1}
},
\nonumber
\\
\label{norm1}
\\
\hat{\mathcal{Z}}^{(2)}_{\mu_{b}\tilde{\mu}_{b}}
&:=& \mathcal{Z}^{(2)}_{\mu_{b}\tilde{\mu}_{b}}/\mathcal{Z}^{(2)}_{\emptyset \emptyset}
\nonumber \\
&=&
\tilde{Z}_{\mu^{t}_{b}}(q,t)\tilde{Z}_{\tilde{\mu}^{t}_{b}}(q,t)t^{\frac{||\mu^{t}_{b}||^2 + ||\tilde{\mu}^{t}_{b}||^2}{2}}
\nonumber \\
&&\times
\prod_{(i,j)\in \mu_{b}}
\frac{
1-Q_{2}Q_{f}t^{-i+\frac{1}{2}}q^{\mu_{b,i}-j+\frac{1}{2}}
}{
1-Q_{f}t^{\tilde{\mu}^{t}_{b,j}-i}q^{\mu_{b,i}-j+1}
}
\prod_{(i,j)\in \tilde{\mu}_{b}}
\frac{
1-Q_{2}t^{-i+\frac{1}{2}}q^{\tilde{\mu}_{b,i}-j+\frac{1}{2}}
}{
1-Q_{f}t^{-\mu^{t}_{b,j}+i-1}q^{-\tilde{\mu}_{b,i}+j}
}.
\label{norm2}
\nonumber \\
\end{eqnarray}
Thus the partition function $\mathcal{Z}_{\text{Local}~\mathcal{B}_{3}}(Q;t,q)$ is as follow:
\begin{eqnarray}
\mathcal{Z}_{\text{Local}~\mathcal{B}_{3}}(Q;t,q)
&=&
\mathcal{Z}^{(1)}_{\emptyset \emptyset}\mathcal{Z}^{(2)}_{\emptyset \emptyset}
\sum_{\mu_{b},\tilde{\mu}_{b}}
(-Q_{b})^{|\mu_{b}|}(-\tilde{Q}_{b})^{|\tilde{\mu}_{b}|}
f^2_{\tilde{\mu}_{b}}(t,q)
\hat{\mathcal{Z}}^{(1)}_{\mu_{b}\tilde{\mu}_{b}}\hat{\mathcal{Z}}^{(2)}_{\mu_{b}\tilde{\mu}_{b}}.
\end{eqnarray}

\subsection{Non-perturbative free energy of the topological string}
Now we define the perturbative free energy of the refined topological string $F(Q;t,q)$ and the unrefined topological string $F_{\text{WS}}(Q;q)$ as follows:
\begin{eqnarray}
F(Q;t,q):=-\text{log}[\mathcal{Z}(Q;t,q)],~F_{\text{WS}}(Q;q)=F(Q;q).
\end{eqnarray}
Then, we define the perturbative parts  of the non-perturbative free energy as
\begin{eqnarray}
F_{\text{WS}}(\mathrm{e}^{-2\pi t_{1}/h +\pi\mathrm{i}},\mathrm{e}^{-2\pi t_{2}/h-\pi\mathrm{i}},\mathrm{e}^{-2\pi t_{b}/h-\pi\mathrm{i}},\mathrm{e}^{-2\pi t_{f}/h};\mathrm{e}^{4\pi^2 \mathrm{i}/\hbar}).
\end{eqnarray}
Note that we redefine the K\"ahler parameters due to the HMO cancellation.
\par
The non-perturbative parts of the non-perturbative free energy are obtained by using the  Nekrasov--Shatashvili limit of the refined topological string \cite{Nekrasov:2009rc},
\begin{eqnarray}
F_{\text{NS}}(Q;q) &:=&\lim_{\epsilon_{2}\to0}\epsilon_{2}F(Q;t,q)
\\
&&(q=\mathrm{e}^{\epsilon_{1}},~t=\mathrm{e}^{-\epsilon_{2}})
\nonumber
\end{eqnarray}
Then, the non-perturbative parts of the free energy $F_{\text{M2}}(\bold{t};\hbar)$ are defined as follow:
\begin{eqnarray}
F_{\text{M2}}(\bold{t};\hbar)
=
\frac{\mathrm{i}}{2\pi}\Bigl[
\bold{t}\cdot \frac{\partial}{\partial \bold{t}} F_{\text{NS}}(\mathrm{e}^{-t_{i}};\mathrm{e}^{\mathrm{i}\hbar})
+
\hbar^2 \frac{\partial}{\partial \hbar}\bigl(F_{\text{NS}}(\mathrm{e}^{-t_{i}};\mathrm{e}^{\mathrm{i}\hbar})/h \bigr)
\Bigr],
\end{eqnarray}
where we define
\begin{eqnarray}
\hbar :=4 \pi^2/g_{s},
\label{hbar and gs}
\end{eqnarray}
and $\bold{t}:=(t_{1},t_{2},t_{b},t_{f})$.
\par
Thus the non-perturbative free energy of the topological string on the local $\mathcal{B}_{3}$ $J_{\text{Local}~\mathcal{B}_{3}}(\bold{t};\hbar)$ is as follow:
\begin{eqnarray}
J_{\text{Local}~\mathcal{B}_{3}}(\bold{t};\hbar)
:=
F_{\text{WS}}(\mathrm{e}^{-2\pi t_{1}/h +\pi\mathrm{i}},\mathrm{e}^{-2\pi t_{2}/h-\pi\mathrm{i}},\mathrm{e}^{-2\pi t_{b}/h-\pi\mathrm{i}},\mathrm{e}^{-2\pi t_{f}/h};\mathrm{e}^{4\pi^2 \mathrm{i}/\hbar})
+
F_{\text{M2}}(\bold{t};\hbar).
\nonumber \\
\end{eqnarray}
In this case, $J_{\text{Local}~\mathcal{B}_{3}}(\bold{t};\hbar)$ is finite for any $g_{s}$ or $\hbar$. For example, when we set $\hbar=2\pi$, then we obtain
\begin{eqnarray}
&&\lim_{\hbar\to2\pi}J_{\text{Local}~\mathcal{B}_{3}}(\bold{t};\hbar)
\nonumber \\
&&
=
-
\frac{2+\pi^2+2t_{1}+t^2_{1}}{8\pi^2}\mathrm{e}^{-t_{1}}
-
\frac{2+\pi^2+2t_{2}+t^2_{2}}{8\pi^2}\mathrm{e}^{-t_{2}}
+
\frac{2+\pi^2 +2t_{b}+t_{b}^2}{8\pi^2}\mathrm{e}^{-t_{b}}
+
\frac{2+2t_{f}+t^2_{f}}{4\pi^2}\mathrm{e}^{-t_{f}}
\nonumber \\
&&
-
\frac{2+\pi^2+2t_{1}+2t_{f} + 2t_{1}t_{f}+t^2_{1}+t^2_{f}}{8\pi^2}\mathrm{e}^{-t_{1}-t_{f}}
-
\frac{2+\pi^2+2t_{2}+2t_{f} + 2t_{2}t_{f}+t^2_{2}+t^2_{f}}{8\pi^2}\mathrm{e}^{-t_{2}-t_{f}}
\nonumber \\
&&
-
\frac{2+9\pi^2+2t_{b}-6t_{f} - 6t_{b}t_{f}+t^2_{b}-3t^2_{f}}{8\pi^2}\mathrm{e}^{-t_{b}-t_{f}}
+\cdots.
\end{eqnarray}
More general discussion for this pole cancellations is written in the reference \cite{Hatsuda:2012dt}.

\section{Geometric transition in non-perturbative topological string}
In this section, we consider the geometric transition.
\par
We first consider the geometric transition from the local $\mathcal{B}_{3}$ to the resolved conifold. In order to know how to set the K\"ahler parameters to occur this geometric transition, we consider the geometric transition in the perturbative topological string in the beginning. After the consideration, we consider the geometric transition in the non-perturbative topological string. Then, we find that the non-perturbative free energy after the geometric transition agrees with the one on the resolved conifold which is obtained in the reference \cite{Hatsuda:2015oaa}. We also find that the K\"ahler parameters are corrected by the non-perturbative effects.
\par
We next consider the geometric transition from the local $\mathcal{B}_{3}$ to the resolved conifold with a toric A-brane. As is the above case, we consider the geometric transition in the perturbative topological string in the beginning. After the consideration, we consider the geometric transition in the non-perturbative topological string. Then, we find that the non-perturbative parts of this free energy after the geometric transition have the same structure as the one in the reference \cite{Marino:2016rsq}.

\subsection{Geometric transition from Local $\mathcal{B}_{3}$ to Resolved conifold}
In this subsection, we consider the geometric transition from the local $\mathcal{B}_{3}$ to the resolved conifold in the closed string.
\begin{figure}[htbp]
\centering
    \includegraphics[width=15cm]{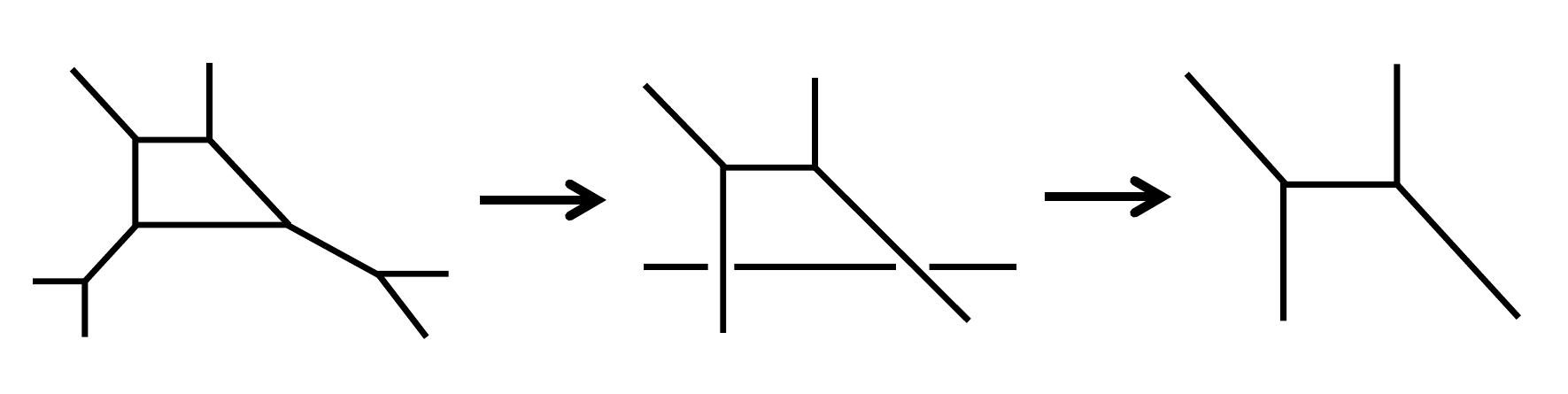}
    \caption{The geometric transition in the closed topological string.}
    \label{geomtransclosed}
\end{figure}

\subsubsection*{Perturbative free energy}
To begin with, we consider how to set the K\"ahler parameters to special values in the refined topological string. According to the reference \cite{Gomis:2006mv}\cite{Gomis:2007kz}\cite{Taki:2010bj}\cite{Dimofte:2010tz}, we set the parameters $Q_{1}$ and $Q_{2}$ as follows:
\begin{eqnarray}
Q_{1}=\sqrt{\frac{t}{q}},~Q_{2}=\sqrt{\frac{q}{t}}.
\end{eqnarray}
Then, the factors in $\hat{\mathcal{Z}}^{(1)}_{\mu_{b}\tilde{\mu}_{b}}$
\begin{eqnarray}
\prod_{(i,j)\in \tilde{\mu}_{b}}
(1-Q_{1}t^{-i+\frac{1}{2}}q^{\tilde{\mu}_{b,i}-j+\frac{1}{2}})
=
\prod_{(i,j)\in \tilde{\mu}_{b}}
(1-t^{-i+1}q^{\tilde{\mu}_{b,i}-j})
\end{eqnarray}
become zero unless the Young diagram $\tilde{\mu}_{b}$ becomes empty. Then, after several cancellation, we obtain
\begin{eqnarray}
\mathcal{Z}_{\text{Local}~\mathcal{B}_{3}}(Q;t,q)
&=&
\prod_{i,j=1}^{\infty}(1-t^{i} q^{j-1} )(1-t^{i-1} q^{j} )
\nonumber \\ 
&&\times
\sum_{\mu_{b}}
(-Q_{b})^{|\mu_{b}|}
\tilde{Z}_{\mu_{b}}(t,q)\tilde{Z}_{\mu^{t}_{b}}(q,t)q^{\frac{||\mu_{b}||^2}{2}}t^{\frac{||\mu^{t}_{b}||^2}{2}}.
\end{eqnarray}
This expression agrees with the partition function of the closed topological string on the resolved conifold except for the factors $\prod_{i,j=1}^{\infty}(1-t^{i} q^{j-1} )(1-t^{i-1} q^{j} )$
\footnote{
This difference is due to the normalization of the topological vertex.}.
Thus, in unrefined case, this geometric transition occurs when we set
\begin{eqnarray}
Q_{1}=Q_{2}=1.
\end{eqnarray}
The same can be said of the perturbative free energy by the relation between the partition function and the free energy.

\par
\subsubsection*{Non-perturbative free energy}
We next consider the non-perturbative free energy. Then, by considering the HMO cancellation mechanism, the K\"ahler parameters of the perturbative parts are replaced as follows:
\begin{eqnarray}
&&t_{1} \rightarrow -2\pi t_{1}/\hbar + \pi\mathrm{i},
\\
&&t_{2} \rightarrow -2\pi t_{2}/\hbar - \pi\mathrm{i},
\\
&&t_{f} \rightarrow -2\pi t_{f}/\hbar - \pi\mathrm{i}.
\end{eqnarray}
 Thus, in order to consider the above geometric transition in the perturbative part, we set the K\"ahler parameters as follows:
\begin{eqnarray}
&&
-\frac{2\pi t_{1}}{\hbar} + \pi\mathrm{i}=0,
\label{closed t1}
\\
&&
-\frac{2\pi t_{2}}{\hbar} - \pi\mathrm{i}=0.
\label{closed t2}
\end{eqnarray}
By using the relation (\ref{hbar and gs}), we obtain
\begin{eqnarray}
&&
t_{1} = \frac{2\pi^2 \mathrm{i}}{g_{s}},
\\
&&
t_{2} = -\frac{2\pi^2 \mathrm{i}}{g_{s}}.
\end{eqnarray}
This means that the K\"ahler parameters are corrected by the non-perturbative parts. Then the non-perturbative parts $F_{\text{M2}}(\bold{t};\hbar)$ become
\begin{eqnarray}
&&F_{\text{M2}}(\bold{t};\hbar)|_{t_{1},t_{2}~\text{fixed}}
\nonumber \\
&&~~~~~~
=
f(\hbar)
+
\frac{
(\hbar/2) \mathrm{cos}[\frac{\hbar}{2}]
+(1+t_{b})\mathrm{sin}[\frac{\hbar}{2}]
}{
4\pi\mathrm{sin}^2[\frac{\hbar}{2}]
}
\mathrm{e}^{-t_{b}}
+
\frac{
\hbar\mathrm{cos}[\hbar]
+(1+2t_{b})\mathrm{sin}[\hbar]
}{
16\pi\mathrm{sin}^2[\hbar]
}
\mathrm{e}^{-2t_{b}}
\nonumber \\
&&~~~~~~~~~~
\cdots,
\end{eqnarray}
where $f(\hbar)$ is the function which is independent of the K\"ahler parameters.\footnote{It would be interesting to consider the meaning of $f(\hbar)$ in terms of the constant map. We would like to thank Sanefumi Moriyama for discussing it.} Thus this function corresponds to the factors $\prod_{i,j=1}^{\infty}(1-t^{i} q^{j-1} )(1-t^{i-1} q^{j} )$. This expression agrees with the one on the resolved conifold which is obtained in the reference \cite{Hatsuda:2015oaa} except for $f(\hbar)$. Therefore we conclude that the geometric transition can be applied, even if we consider the non-perturbative topological string.

\subsection{Geometric transition from  Local $\mathcal{B}_{3}$ to Resolved conifold with a Toric A-brane}
In this subsection, we consider the geometric transition from the closed topological string on the local $\mathcal{B}_{3}$ to the open topological string on the resolved conifold with a toric A-brane.
\begin{figure}[htbp]
\centering
    \includegraphics[width=15cm]{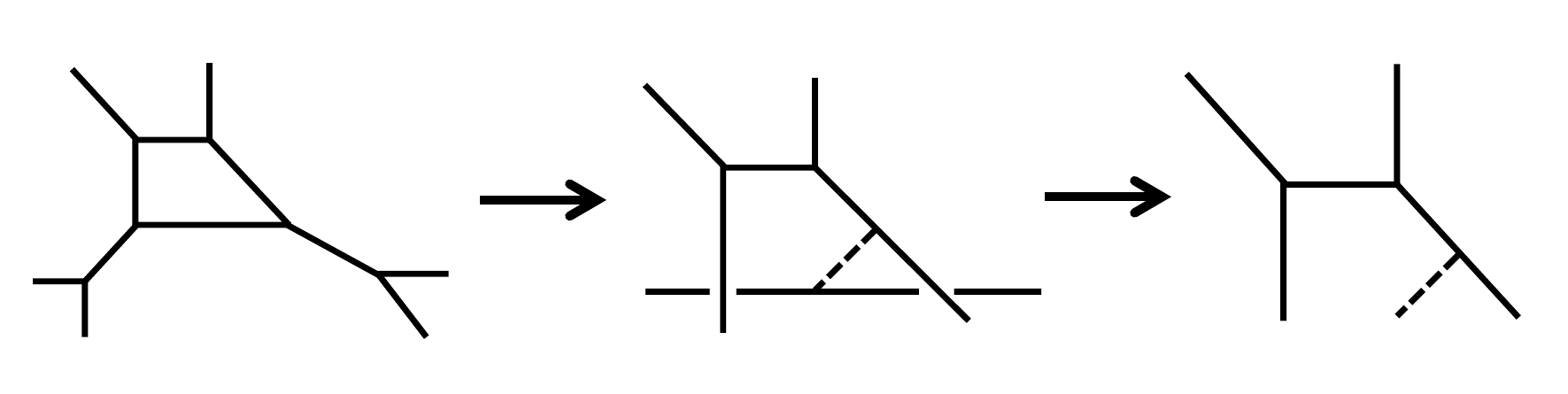}
    \caption{The geometric transition from the closed topological string to the open topological string. }
    \label{geomtransopen}
\end{figure}

\subsubsection*{Perturbative free energy}

We consider again the refined topological string in the beginning. According to the reference \cite{Gomis:2006mv}\cite{Gomis:2007kz}\cite{Taki:2010bj}\cite{Dimofte:2010tz}, we set the parameters $Q_{1}$ and $Q_{2}$ as follows:
\begin{eqnarray}
Q_{1}=\sqrt{\frac{t}{q}},~Q_{2}=t\sqrt{\frac{q}{t}}
\end{eqnarray}
Then, as we said in previous section, the Young diagram $\tilde{\mu}_{b}$ becomes empty. Thus, by using the expressions (\ref{norm1}) and (\ref{norm2}), we obtain
\begin{eqnarray}
\mathcal{Z}_{\text{Local}~\mathcal{B}_{3}}(Q;t,q)
&=&
\prod_{i,j=1}(1-t^{i}q^{j})(1-t^{i}q^{j-1})
\nonumber \\
&&\times
\sum_{\mu_{b}}(-Q_{b})^{|\mu_{b}|}
\tilde{Z}_{\mu_{b}}(t,q)\tilde{Z}_{\mu^{t}_{b}}(q,t)q^{\frac{||\mu_{b}||^2}{2}}t^{\frac{||\mu^{t}_{b}||^2}{2}}
\prod_{j=1}^{\infty}
\frac{
1
}{
1-Q_{f}t^{-\mu_{b,j}^{t}}q^{j}
}.
\nonumber \\
\label{Zclosedopen}
\end{eqnarray}
\par
In order to check the consistency, we calculate the partition function of the open topological string on the resolved conifold with a toric A-brane. Its web diagram is shown in Fig.\ref{geomtransopen}. Then we write the partition function from the web diagram as follow:
\begin{eqnarray}
\mathcal{Z}_{\text{open}}
=
\sum_{\text{all indices}}
(-Q_{b})^{|\mu|}C_{\emptyset \emptyset \mu}(t,q)C_{\mu' \emptyset \mu^t}(q,t)\mathrm{Tr}_{{\mu'^{t}}}V,
\end{eqnarray}
where $V$ is the holonomy matrix. Since there is a single A-brane, the matrix $V$ is the one by one matrix,
\begin{eqnarray}
V=\text{diag}(z).
\end{eqnarray}
Then, we obtain
\begin{eqnarray}
\mathcal{Z}_{\text{open}}
=
\sum_{\mu}(-Q_{b})^{|\mu|}
\tilde{Z}_{\mu}(t,q)\tilde{Z}_{\mu^{t}}(q,t)q^{\frac{||\mu||^2}{2}}t^{\frac{||\mu^{t}||^2}{2}}
\prod_{j=1}^{\infty}
\frac{
1
}{
1-zt^{-\mu_{j}^{t}+\frac{1}{2}}q^{j-1}
}.
\label{eq:Zopen}
\end{eqnarray}
Then, (\ref{Zclosedopen}) agrees with (\ref{eq:Zopen}) under the relation between $z$ and $Q_{f}$,
\begin{eqnarray}
z=t^{-\frac{1}{2}}qQ_{f},
\label{z and Qf}
\end{eqnarray}
except for the factors $\prod_{i,j=1}(1-t^{i}q^{j})(1-t^{i}q^{j-1})$. Thus, in the perturbative topological string, the geometric transition occurs when we set
\begin{eqnarray}
Q_{1}=1,~Q_{2}=q.
\end{eqnarray}

\subsubsection*{Non-perturbative free energy}
We now consider the non-perturbative free energy of the open topological string. According to the above discussion, we set the K\"ahler parameters for the geometric transition in the perturbative parts:
\begin{eqnarray}
&&
t_{1} = \frac{2\pi^2  \mathrm{i}}{g_{s}} ,
\\
&&
t_{2} = -\frac{2\pi^2 \mathrm{i}}{g_{s}}- 2\pi \mathrm{i}.
\end{eqnarray}
Then, we obtain the non-perturbative effects $F_{\text{M2}}(\bold{t};\hbar)$
\begin{eqnarray}
&&
F_{\text{M2}}(\bold{t};\hbar)|_{t_{1},t_{2}~\text{fixed}}
\nonumber \\
&&~~~~
=
g(\hbar)
+
\frac{
(\hbar/2) \mathrm{cos}[\frac{\hbar}{2}]
+(1+t_{b})\mathrm{sin}[\frac{\hbar}{2}]
}{
4\pi\mathrm{sin}^2[\frac{\hbar}{2}]
}
\mathrm{e}^{-t_{b}}
+
\frac{
\hbar\mathrm{cos}[\hbar]
+(1+2t_{b})\mathrm{sin}[\hbar]
}{
16\pi\mathrm{sin}^2[\hbar]
}
\mathrm{e}^{-2t_{b}}
\nonumber \\
&&~~~~~~
+
\frac{
\mathrm{i}\mathrm{e}^{\mathrm{i}\hbar/2}
}{
2\mathrm{sin}[\frac{\hbar}{2}]
}
\mathrm{e}^{-t_{f}}
+
\frac{
\mathrm{i}\mathrm{e}^{\mathrm{i}\hbar}
}{
2\mathrm{sin}[\frac{\hbar}{2}]
}
\mathrm{e}^{-t_{b}-t_{f}}
+
\biggl(
\frac{
\mathrm{i}\mathrm{e}^{5\mathrm{i}\hbar/2}
}{
2\mathrm{sin}[\hbar]
}
+
\frac{
\mathrm{i}\mathrm{e}^{3\mathrm{i}\hbar/2}
}{
2\mathrm{sin}[\hbar]
}
\biggr)
\mathrm{e}^{-t_{b}-2t_{f}}
\nonumber \\
&&~~~~~
-
\biggl(
\frac{
\mathrm{i}\mathrm{e}^{2\mathrm{i}\hbar}
}{
4\mathrm{sin}[\hbar]
}
+
\frac{
\mathrm{i}\mathrm{e}^{3\mathrm{i}\hbar}
}{
2\mathrm{sin}[\hbar]
}
\biggr)
\mathrm{e}^{-2t_{b}-2t_{f}}
+
\cdots,
\end{eqnarray}
where $g(\hbar)$ is the function which is independent of the K\"ahler parameters. Moreover, according to the relation (\ref{z and Qf}),  we set the correspondence between $Q_{f}$ and $z$ as follow,
\begin{eqnarray}
-\frac{2\pi}{\hbar}t_{f} +\frac{2\pi^2 \mathrm{i}}{\hbar} = -\frac{2\pi}{\hbar}x
\Leftrightarrow
t_{f}=x+\pi\mathrm{i},
\end{eqnarray}
where we define $z=\mathrm{e}^{-x}$. Thus, we obtain
\begin{eqnarray}
&&
F_{\text{M2}}(\bold{t};\hbar)|_{t_{1},t_{2}~\text{fixed}}
\nonumber \\
&&~~~~
=
g(\hbar)
+
\frac{
(\hbar/2) \mathrm{cos}[\frac{\hbar}{2}]
+(1+t_{b})\mathrm{sin}[\frac{\hbar}{2}]
}{
4\pi\mathrm{sin}^2[\frac{\hbar}{2}]
}
\mathrm{e}^{-t_{b}}
+
\frac{
\hbar\mathrm{cos}[\hbar]
+(1+2t_{b})\mathrm{sin}[\hbar]
}{
16\pi\mathrm{sin}^2[\hbar]
}
\mathrm{e}^{-2t_{b}}
\nonumber \\
&&~~~~~~
+
\frac{
\mathrm{i}\mathrm{e}^{\mathrm{i}\hbar/2}
}{
2\mathrm{sin}[\hbar]
}
(-\mathrm{e}^{-x})
+
\frac{
\mathrm{i}\mathrm{e}^{\hbar}
}{
2\mathrm{sin}[\frac{\hbar}{2}]
}
\mathrm{e}^{-t_{b}}(-\mathrm{e}^{-x})
+
\biggl(
\frac{
\mathrm{i}\mathrm{e}^{5\mathrm{i}\hbar/2}
}{
2\mathrm{sin}[\hbar]
}
+
\frac{
\mathrm{i}\mathrm{e}^{3\mathrm{i}\hbar/2}
}{
2\mathrm{sin}[\hbar]
}
\biggr)
\mathrm{e}^{-t_{b}}
(-\mathrm{e}^{-x})^2
\nonumber \\
&&~~~~~~
-
\biggl(
\frac{
\mathrm{i}\mathrm{e}^{2\mathrm{i}\hbar}
}{
4\mathrm{sin}[\hbar]
}
+
\frac{
\mathrm{i}\mathrm{e}^{3\mathrm{i}\hbar}
}{
2\mathrm{sin}[\hbar]
}
\biggr)
\mathrm{e}^{-2t_{b}}
(-\mathrm{e}^{-x})^2
+
\cdots.
\end{eqnarray}
Then, we find that the non-perturbative parts of the open topological string have the same structure as the one in the reference \cite{Marino:2016rsq}. We check this up to second order of the K\"ahler parameters.
\par
Thus, we obtain the non-perturbative free energy of the open topological string on the resolved conifold with a toric A-brane,
\begin{eqnarray}
J_{X}^{\text{open}}(\bold{t};\hbar)
:=
J_{X}(\bold{t};\hbar)\Bigl|_{t_{1}=\frac{2\pi^2 \mathrm{i}}{g_{s}},~t_{2}=-\frac{2\pi^2 \mathrm{i}}{g_{s}}-2\pi \mathrm{i}}.
\end{eqnarray}
This free energy is finite for any $g_{s}$ or $\hbar$ by the HMO cancellation mechanism. For example, when we set $\hbar=2 \pi$, the free energy $J_{X}^{\text{open}}(\bold{t};\hbar)$ becomes
\begin{eqnarray}
&&\lim_{\hbar\to2\pi}J_{X}^{\text{open}}(\bold{t};\hbar)
\nonumber \\
&&~~=
\frac{
-9+24\pi^2+10\pi\mathrm{i}
}{
16\pi^2
}
+
\frac{
2\pi+\mathrm{i}(1+t_{f})
}{
2\pi
}
\mathrm{e}^{-t_{f}}
+
\frac{
4\pi + \pi\mathrm{i}(1+2t_{f})
}{
8\pi^2
}
\mathrm{e}^{-2t_{f}}
\nonumber \\
&&~~~~
+
\frac{
2+\pi^2+2t_{b}-t_{b}^{2}
}{
8\pi^2
}
\mathrm{e}^{-t_{b}}
-
\frac{
3\pi+\mathrm{i}(1+t_{b}+t_{f})
}{
2\pi
}
\mathrm{e}^{-t_{b}-t_{f}}
-
\frac{
1+2\pi^2 +2t_{b}(1+t_{b})
}{
32\pi^2
}
\mathrm{e}^{-2t_{b}}
\nonumber \\
&&~~~~
+\cdots,
\end{eqnarray}
where $t_{f}=x-\pi\mathrm{i}$.


\section{Summary and Future work}
In this paper, we have considered the geometric transition in the non-perturbative topological string in two cases: One is the geometric transition from the local $\mathcal{B}_{3}$ to the resolved conifold. The other is the geometric transition from the local $\mathcal{B}_{3}$ to the resolved conifold with a toric A-brane. Then we have found that the geometric transition can be applied, even if the non-perturbative effects are included. We also find that the K\"ahler parameters are corrected by the non-perturbative effects. In the open topological string, the non-perturbative free energy which we have obtained has had the same structure as the one in the reference \cite{Marino:2016rsq}.
\par
We have various future work. First of all,  considering the general formula of the non-perturbative open topological string would be interesting. Its structure would be similar to the free energy of the closed topological string which is derived in the reference \cite{Hatsuda:2013oxa}. 
\par
In this paper, we have considered the open topological string with a toric A-brane. Then we want to consider the open topological string in the presence of $N$ A-branes. We would be able to use the geometric transition to obtain this free energy. In this case, we set the K\"ahler parameters as follows,
\begin{eqnarray}
&&
t_{1} = \frac{2\pi^2  \mathrm{i}}{g_{s}} ,
\\
&&
t_{2} = -\frac{2\pi^2 \mathrm{i}}{g_{s}}- 2N\pi \mathrm{i},
\\
&&-\frac{2\pi}{\hbar}t_{f}+\frac{4\pi^2 \mathrm{i}}{\hbar}(j-\frac{1}{2})
=-\frac{2\pi}{\hbar}x_{j},
\\
&&~~~~~~~~~~~~~~~~~~~~~~~~~~~~~~~~~~~(j=1,2,...,N)\nonumber
\end{eqnarray}
where the variables $x_{j}$ correspond to the positions of the A-branes. The justification of this parameter choices would be important.
\par
In terms of the mirror curve, we consider the mirror curve of genus 1 in this paper. Then, naively we can derive the mirror curve of genus 0 by using the geometric transition. On the other hand, according to the reference \cite{Grassi:2014zfa}, the quantization of mirror curve relates to the free energy of topological string on the toric Calabi-Yau manifold associated with the mirror curve. However there is a crucial problem. For the quantization of mirror curve of genus 0, the expectation value of the trace class operator which is defined by the quantization of the mirror curve diverges since the spectrum of this operator is continuous. Then, by using our result and the mirror symmetry \cite{Chiang:1999tz}\cite{Hori:2000kt}, we might obtain the finite expectation value for the mirror curve of genus 0 by setting some parameters in the expectation value of the mirror curve of genus 1 whose expectation value is finite.
\par
The above discussion about the mirror curve can be applied to the open string. Then, the free energy of the open topological string might be able to relate to the quantization of mirror curve. It would be also interesting to consider this.

\section*{Acknowledgement}
I would like to thank Satoshi Yamaguchi and Sanefumi Moriyama for discussions and comments. I also would like to thank Taro Kimura for discussions.

\clearpage

\appendix

\section{Definitions and some formulae}
In this appendix, we summarize the definitions and the formulae which we use in this paper.
\begin{itemize}
\item The refined topological vertex
\begin{eqnarray}
C_{\lambda \mu \nu}(t,q) &=& t^{-\frac{||\mu^{t}||^{2}}{2}}q^{\frac{||\mu||^2 + ||\nu||^{2}}{2}} \tilde{Z}_{\nu}(t,q)\nonumber \\
&& \times \sum_{\eta}\Bigl(\frac{q}{t}\Bigr)^{\frac{|\eta| + |\lambda| - |\mu|}{2}}  s_{\lambda^{t}/\eta}(t^{-\rho}q^{-\nu})s_{\mu/\eta}(t^{-\nu^{t}}q^{-\rho}), \\
\tilde{Z}_{\nu}(t,q) &=& \prod_{(i,j) \in \nu}(1-q^{\nu_{i}-j}t^{\nu_{j}^{t} -i +1})^{-1}.
\nonumber
\\ 
&&
\Bigl(
|\mu|:=\sum_{i=1}^{l(\mu)}\mu_{i},~||\mu||^{2}:=\sum_{i=1}^{l(\mu)}\mu^{2}_{i}
\Bigr)
\nonumber
\end{eqnarray}
\item The gluing factors
\begin{eqnarray}
f_{\mu} (t,q) = (-1)^{|\mu|}q^{-\frac{||\mu||^2}{2}}t^{\frac{||\mu^{t}||^2}{2}},~~
\tilde{f}_{\mu} (t,q) = (-1)^{|\mu|}\Bigl(\frac{t}{q}\Bigr)^{\frac{|\mu|}{2}}q^{-\frac{||\mu||^2}{2}}t^{\frac{||\mu^{t}||^2}{2}},
\end{eqnarray}
\end{itemize}
where $s_{\mu/\eta}(x_{1},x_{2},...)$ is the skew Schur function. We also define the Young diagram $\mu$ as following figure \ref{Young}.
\begin{figure}[htbp]
\centering
    \includegraphics[width=15cm]{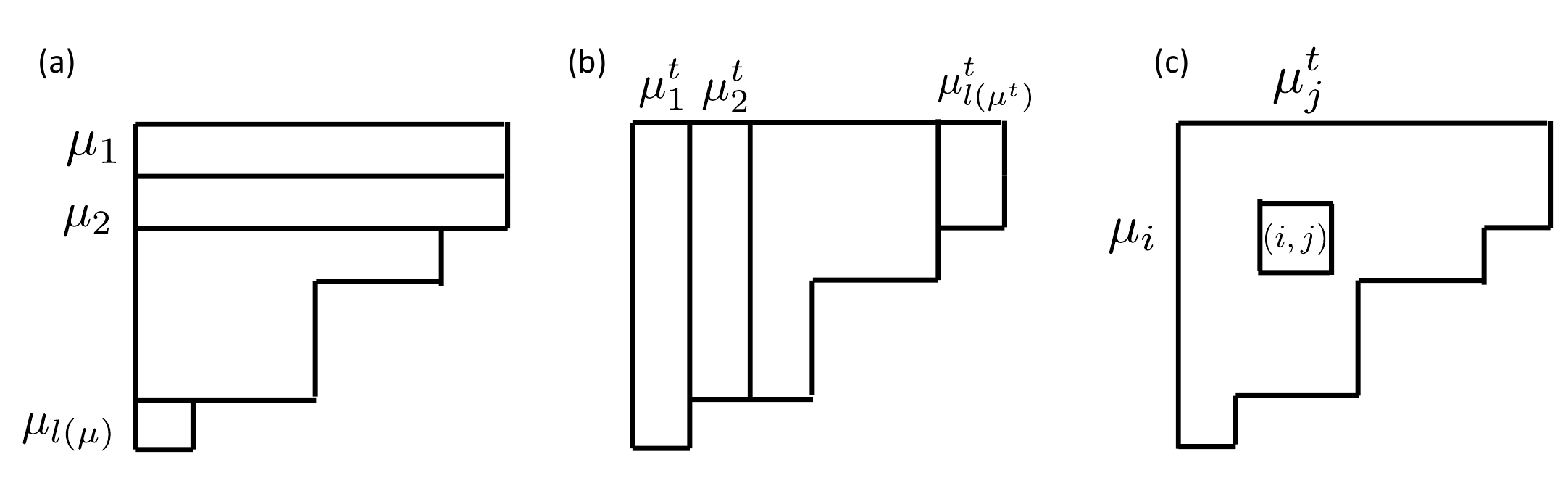}
    \caption{The Young diagram.  We define $\mu_{i}$ as Fig. (a), $\mu^{t}_{j}$ as Fig. (b), and the coordinate $(i,j)$ as Fig. (c). }
    \label{Young}
\end{figure}\\
When we set $t=q$, the refined topological vertex becomes the unrefined topological vertex.
\par
We also write some useful formulae.
\begin{itemize}
\item Some formulas about Schur polynomial
\begin{eqnarray}
s_{\lambda/\mu}(\alpha \bold{x}) &=& \alpha^{|\lambda|-|\mu|}s_{\lambda/\mu}(\bold{x}) \\
\sum_{\eta}s_{\eta/\lambda}(\bold{x})s_{\eta/\mu}(\bold{y}) &=& \prod_{i,j=1}^{\infty}(1-x_{i}y_{j})^{-1}\sum_{\tau}s_{\mu/\tau}(\bold{x})s_{\lambda/\tau}(\bold{y}) \\
\sum_{\eta}s_{\eta^{t}/\lambda}(\bold{x})s_{\eta/\mu}(\bold{y}) &=& \prod_{i,j=1}^{\infty}(1+x_{i}y_{j})\sum_{\tau}s_{\mu^{t}/\tau}(\bold{x})s_{\lambda^{t}/\tau^{t}}(\bold{y})
\end{eqnarray}
\item Normalization
\begin{eqnarray}
\prod_{i,j=1}^{\infty} \frac{1-Qq^{\nu_{i}-j}t^{\mu_{j}^{t}-i+1}}{1-Qq^{-j}t^{-i+1}} &=& \prod_{(i,j) \in \nu}(1-Qq^{\nu_{i}-j}t^{\mu_{j}^{t}-i+1})\prod_{(i,j) \in \mu}(1-Qq^{-\mu_{i}+j-1}t^{-\nu_{j}^{t}+i}) ~~~~~~~~~~~~\\
\prod_{i,j=1}^{\infty} \frac{1-Qt^{\nu_{j}^{t}-i+\frac{1}{2}}q^{-j+\frac{1}{2}}}{1-Qt^{-i+\frac{1}{2}}q^{-j+\frac{1}{2}}} &=& \prod_{(i,j) \in \nu}(1-Qq^{-j+\frac{1}{2}}t^{i-\frac{1}{2}}) \\
\prod_{i,j=1}^{\infty} \frac{1-Qq^{\nu_{i}-j+\frac{1}{2}}t^{-i+\frac{1}{2}}}{1-Qq^{-j+\frac{1}{2}}t^{-i+\frac{1}{2}}} &=&\prod_{(i,j) \in \nu}(1-Qq^{j-\frac{1}{2}}t^{-i+\frac{1}{2}})
\end{eqnarray}
\end{itemize}


\section{Topological string on Local $\mathcal{B}_{3}$}

\subsection{Partition function of Refined topological string}
We write the partition function of the refined topological string on the local $\mathcal{B}_{3}$ for some orders explicitly,
\begin{eqnarray}
&&\mathcal{Z}_{\text{Local}~\mathcal{B}_{3}}(Q;t,q)
\nonumber \\
&&=
\prod_{i,j=1}^{\infty}
\frac{
(1-Q_{1}t^{i-\frac{1}{2}}q^{j-\frac{1}{2}})(1-Q_{2}t^{i-\frac{1}{2}}q^{j-\frac{1}{2}})
(1-Q_{1}Q_{f}t^{i-\frac{1}{2}}q^{j-\frac{1}{2}})(1-Q_{1}Q_{f}t^{i-\frac{1}{2}}q^{j-\frac{1}{2}})
}{
(1-Q_{f}t^{i-1}q^{j})(1-Q_{f}t^{i}q^{j-1})
}
\nonumber \\
&&~~~~
\times
\biggl\{
1-
\biggl[
\frac{
(tq)^{\frac{1}{2}}(1-Q_{1}Q_{f}t^{-\frac{1}{2}}q^{\frac{1}{2}})(1-Q_{2}Q_{f}t^{-\frac{1}{2}}q^{\frac{1}{2}})
}{
(1-t)(1-q)(1-Q_{f})(1-Q_{f}t^{-1}q)
}
+
\frac{
t^{\frac{3}{2}}q^{-\frac{1}{2}}(1-Q_{1}t^{-\frac{1}{2}}q^{\frac{1}{2}})(1-Q_{2}t^{-\frac{1}{2}}q^{\frac{1}{2}})
}{
(1-t)(1-q)(1-Q_{f})(1-Q_{f}tq^{-1})
}
Q_{f}
\biggr]
Q_{b}
\nonumber \\
&&~~~~
+
\biggl[
\biggl(
\frac{
t^{2}q
(1-Q_{1}Q_{f}t^{-\frac{1}{2}}q^{\frac{1}{2}})(1-Q_{1}Q_{f}t^{-\frac{3}{2}}q^{\frac{1}{2}})
(1-Q_{2}Q_{f}t^{-\frac{1}{2}}q^{\frac{1}{2}})(1-Q_{2}Q_{f}t^{-\frac{3}{2}}q^{\frac{1}{2}})
}{
(1-t)(1-t^2)(1-tq)(1-q)(1-Q_{f})(1-Q_{f}t^{-1})(1-Q_{f}t^{-1}q)(1-Q_{f}t^{-2}q)
}
\nonumber \\
&&~~~~
+
\frac{
tq^{2}
(1-Q_{1}Q_{f}t^{-\frac{1}{2}}q^{\frac{1}{2}})(1-Q_{1}Q_{f}t^{-\frac{1}{2}}q^{\frac{3}{2}})
(1-Q_{2}Q_{f}t^{-\frac{1}{2}}q^{\frac{1}{2}})(1-Q_{2}Q_{f}t^{-\frac{1}{2}}q^{\frac{3}{2}})
}{
(1-q)(1-q^2)(1-tq)(1-t)(1-Q_{f})(1-Q_{f}q)(1-Q_{f}qt^{-1})(1-Q_{f}q^{2}t^{-1})
}
\biggr)
\nonumber \\
&&~~~~
+
\frac{
t^2
(1-Q_{1}t^{-\frac{1}{2}}q^{\frac{1}{2}})(1-Q_{1}Q_{f}t^{-\frac{1}{2}}q^{\frac{1}{2}})
(1-Q_{2}t^{-\frac{1}{2}}q^{\frac{1}{2}})(1-Q_{2}Q_{f}t^{-\frac{1}{2}}q^{\frac{1}{2}})
}{
(1-t)^{2}(1-q)^{2}(1-Q_{f}t)(1-Q_{f}q)(1-Q_{f}t^{-1})(1-Q_{f}q^{-1})
}
Q_{f}
+
\nonumber \\
&&~~~~
\biggl(
\frac{
t^{6}q^{-1}
(1-Q_{1}t^{-\frac{1}{2}}q^{\frac{1}{2}})(1-Q_{1}t^{-\frac{3}{2}}q^{\frac{1}{2}})
(1-Q_{2}t^{-\frac{1}{2}}q^{\frac{1}{2}})(1-Q_{2}t^{-\frac{3}{2}}q^{\frac{1}{2}})
}{
(1-t)(1-t^2)(1-tq)(1-q)(1-Q_{f})(1-Q_{f}t)(1-Q_{f}tq^{-1})(1-Q_{f}t^{2}q^{-1})
}
+
\nonumber \\
&&~~~~
\frac{
t^{3}q^{-2}
(1-Q_{1}t^{-\frac{1}{2}}q^{\frac{1}{2}})(1-Q_{1}t^{-\frac{1}{2}}q^{\frac{3}{2}})
(1-Q_{2}t^{-\frac{1}{2}}q^{\frac{1}{2}})(1-Q_{2}t^{-\frac{1}{2}}q^{\frac{3}{2}})
}{
(1-q)(1-q^2)(1-tq)(1-t)(1-Q_{f})(1-Q_{f}q^{-1})(1-Q_{f}tq^{-1})(1-Q_{f}tq^{-2})
}
\biggr)
Q^{2}_{f}
\biggr]
Q^{2}_{b}
+
\cdots
\biggr\}.
\nonumber \\
\end{eqnarray}
In this paper, we use this expression.

\subsection{Free energy of topological string}
\subsection*{Free energy of topological string}
The free energy of the unrefined closed topological string on the local $\mathcal{B}_{3}$ is as follow:
\begin{eqnarray}
F_{\text{WS}}(Q;q)
&=&
\Biggl(
\sum_{m=1}^{\infty}
\frac{
Q^{m}_{1}+Q^{m}_{2}+(Q_{1}Q_{f})^{m}+(Q_{2}Q_{f})^{m}-2Q_{f}^{m}
}{
m(q^{\frac{m}{2}}-q^{-\frac{m}{2}})^2
}
\Biggr)
\nonumber \\
&&~~
+
\Biggl(
\frac{
1
}{
(q^{\frac{1}{2}}-q^{-\frac{1}{2}})^2
}
+
\frac{
3
}{
(q^{\frac{1}{2}}-q^{-\frac{1}{2}})^2
}
Q_{f}
+
\frac{
5
}{
(q^{\frac{1}{2}}-q^{-\frac{1}{2}})^2
}
Q^2_{f}
-
\frac{
2
}{
(q^{\frac{1}{2}}-q^{-\frac{1}{2}})^2
}
Q_{2}Q_{f}
\nonumber \\
&&~~
-
\frac{
4
}{
(q^{\frac{1}{2}}-q^{-\frac{1}{2}})^2
}
Q_{2}Q^2_{f}
-
\frac{
2
}{
(q^{\frac{1}{2}}-q^{-\frac{1}{2}})^2
}
Q_{1}Q_{f}
-
\frac{
4
}{
(q^{\frac{1}{2}}-q^{-\frac{1}{2}})^2
}
Q_{1}Q^2_{f}
+
\frac{
1
}{
(q^{\frac{1}{2}}-q^{-\frac{1}{2}})^2
}
Q_{1}Q_{2}Q_{f}
\nonumber \\
&&~~
+
\frac{
3
}{
(q^{\frac{1}{2}}-q^{-\frac{1}{2}})^2
}
Q_{1}Q_{2}Q^2_{f}
\Biggr)Q_{b}
\nonumber \\
&&~~
+
\Biggl(
\frac{
1
}{
2(q-q^{-1})^2
}
-
\biggl(
\frac{
6
}{
(q^{\frac{1}{2}}-q^{-\frac{1}{2}})^2
}
-
\frac{
3
}{
2(q-q^{-1})^2
}
\biggr)
Q^2_{f}
+
\frac{
5
}{
(q^{\frac{1}{2}}-q^{-\frac{1}{2}})^2
}
Q_{2}Q^{2}_{f}
\nonumber \\
&&~~
-
\frac{
2
}{
2(q-q^{-1})^2
}
Q^2_{2}Q^2_{f}
+
\frac{
5
}{
(q^{\frac{1}{2}}-q^{-\frac{1}{2}})^2
}
Q_{1}Q^{2}_{f}
-
\frac{
2
}{
2(q-q^{-1})^2
}
Q_{1}^{2}Q_{f}^{2}
-
\frac{
4
}{
(q^{\frac{1}{2}}-q^{-\frac{1}{2}})^2
}
Q_{1}Q_{2}Q^{2}_{f}
\Biggr)
Q^2_{b}
\nonumber \\
&&~~
+\cdots.
\end{eqnarray}

\subsection*{Free energy of topological string in NS limit}
The NS limit for the free energy of the closed topological string on the local $\mathcal{B}_{3}$ is as follow:
\begin{eqnarray}
F_{\text{NS}}(Q;q)
&=&
\Biggl(
\sum_{m=1}^{\infty}
\frac{
Q^{m}_{1}+Q^{m}_{2}+(Q_{1}Q_{f})^{m}+(Q_{2}Q_{f})^{m}
}{
m^2(q^{\frac{m}{2}}-q^{-\frac{m}{2}})
}
-
\sum_{m=1}^{\infty}
\frac{
Q_{f}^{m}(q^{m}-q^{-m})
}{
m^2(q^{\frac{m}{2}}-q^{-\frac{m}{2}})^2
}
\Biggr)
\nonumber \\
&&~~
+
\Biggl(
\frac{
1
}{
q^{\frac{1}{2}}-q^{-\frac{1}{2}}
}
+
\frac{
q^{\frac{3}{2}}-q^{-\frac{3}{2}}
}{
(q^{\frac{1}{2}}-q^{-\frac{1}{2}})^2
}
Q_{f}
+
\frac{
q^{\frac{5}{2}}-q^{-\frac{5}{2}}
}{
(q^{\frac{1}{2}}-q^{-\frac{1}{2}})^2
}
Q^2_{f}
-
\frac{
q-q^{-1}
}{
(q^{\frac{1}{2}}-q^{-\frac{1}{2}})^2
}
Q_{1}Q_{f}
\nonumber \\
&&~~
-
\frac{
q-q^{-1}
}{
(q^{\frac{1}{2}}-q^{-\frac{1}{2}})^2
}
Q_{2}Q_{f}
-
\frac{
q^{2}-q^{-2}
}{
(q^{\frac{1}{2}}-q^{-\frac{1}{2}})^2
}
Q_{1}Q^{2}_{f}
-
\frac{
q^{2}-q^{-2}
}{
(q^{\frac{1}{2}}-q^{-\frac{1}{2}})^2
}
Q_{2}Q^{2}_{f}
+
\frac{
1
}{
q^{\frac{1}{2}}-q^{-\frac{1}{2}}
}
Q_{1}Q_{2}Q_{f}
\nonumber \\
&&~~
+
\frac{
q^{\frac{3}{2}}-q^{-\frac{3}{2}}
}{
(q^{\frac{1}{2}}-q^{-\frac{1}{2}})^2
}
Q_{1}Q_{2}Q^{2}_{f}
\Biggr)
Q_{b}
\nonumber \\
&&~~
+
\Biggl(
\frac{
1
}{
4(q-q^{-1})
}
+
\biggl(
\frac{
q^{2}-q^{-2}
}{
(q-q^{-1})^2
}
+
\frac{
7(q^{3}-q^{-3})
}{
4(q-q^{-1})^2
}
+
\frac{
q^{4}-q^{-4}
}{
(q-q^{-1})^2
}
\biggr)
Q^{2}_{f}
+
\frac{
q^{\frac{5}{2}}-q^{-\frac{5}{2}}
}{
(q^{\frac{1}{2}}-q^{-\frac{1}{2}})^2
}
Q_{1}Q^{2}_{f}
\nonumber \\
&&~~
+
\frac{
q^{\frac{5}{2}}-q^{-\frac{5}{2}}
}{
(q^{\frac{1}{2}}-q^{-\frac{1}{2}})^2
}
Q_{2}Q^{2}_{f}
-
\frac{
q^{2}-q^{-2}
}{
4(q-q^{-1})^2
}
Q^{2}_{1}Q^{2}_{f}
-
\frac{
q^{2}-q^{-2}
}{
4(q-q^{-1})^2
}
Q^{2}_{2}Q^{2}_{f}
-
\frac{
q^{2}-q^{-2}
}{
(q^{\frac{1}{2}}-q^{-\frac{1}{2}})^2
}
Q_{1}Q_{2}Q^{2}_{f}
\nonumber \\
&&~~
+
\frac{
1
}{
4(q-q^{-1})
}
Q^{2}_{1}Q^{2}_{2}Q^{2}_{f}
\Biggr)
Q^{2}_{b}
+\cdots.
\end{eqnarray}

\section{Proof of geometric transition from Local $\mathcal{B}_{3}$ to Resolved conifold}
In this section, we show that the non-perturbative parts of the free energy of the local $\mathcal{B}_{3}$ become the one of the resolved conifold after setting the K\"ahler parameters to (\ref{closed t1}) and (\ref{closed t2}).
\par
We first consider some derivatives. We define
\begin{eqnarray}
W^{(1)}_{\mu_{b}\tilde{\mu}_{b}}(Q;t,q)
&:=&
\prod_{(i,j)\in \mu_{b}}
\frac{
1-Q_{1}Q_{f}t^{-i+\frac{1}{2}}q^{\mu_{b,i}-j+\frac{1}{2}}
}{
1-Q_{f}t^{\tilde{\mu}^{t}_{b,j}-i+1}q^{\mu_{b,i}-j}
}
\prod_{(i,j)\in \tilde{\mu}_{b}}
\frac{
1-Q_{1}t^{-i+\frac{1}{2}}q^{\tilde{\mu}_{b,i}-j+\frac{1}{2}}
}{
1-Q_{f}t^{-\mu^{t}_{b,j}+i}q^{-\tilde{\mu}_{b,i}+j-1}
},
\nonumber \\ \\
W^{(2)}_{\mu_{b}\tilde{\mu}_{b}}(Q;t,q)
&:=&
\prod_{(i,j)\in \mu_{b}}
\frac{
1-Q_{2}Q_{f}t^{-i+\frac{1}{2}}q^{\mu_{b,i}-j+\frac{1}{2}}
}{
1-Q_{f}t^{\tilde{\mu}^{t}_{b,j}-i}q^{\mu_{b,i}-j+1}
}
\prod_{(i,j)\in \tilde{\mu}_{b}}
\frac{
1-Q_{2}t^{-i+\frac{1}{2}}q^{\tilde{\mu}_{b,i}-j+\frac{1}{2}}
}{
1-Q_{f}t^{-\mu^{t}_{b,j}+i-1}q^{-\tilde{\mu}_{b,i}+j}
}.
\nonumber \\
\end{eqnarray}
These expressions are the factors of $\hat{\mathcal{Z}}_{\mu_{b}\tilde{\mu}_{b}}^{(1)}$ and $\hat{\mathcal{Z}}_{\mu_{b}\tilde{\mu}_{b}}^{(2)}$. The Nekrasov--Shatashvili limit does not affect $W^{(1)}_{\mu_{b}\tilde{\mu}_{b}}(Q;t,q)$, $W^{(2)}_{\mu_{b}\tilde{\mu}_{b}}(Q;t,q)$ since they do not have the pole. Then, by performing the derivatives, we obtain
\begin{eqnarray}
&&\Bigl[\partial_{t_{1}}[
\lim_{\epsilon_{2}\to0}
W^{(1)}_{\mu_{b}\tilde{\mu}_{b}}(Q;t,q)W^{(2)}_{\mu_{b}\tilde{\mu}_{b}}(Q;t,q)]\Bigr]
|_{t_{1},~t_{2}~\text{fixed}}
\nonumber \\
&&=
\sum_{\mu_{b}}
\frac{
Q_{f}q^{\mu_{b,i}-j}
}{
1-Q_{f}q^{\mu_{b,i}-j}
}
+
\frac{
1
}{
1-Q_{f}q^{-1}
}
\prod_{(i,j)\in \tilde{\mu}_{b},\tilde{\mu}_{b,i}-j\neq0}
\frac{
1-q^{\tilde{\mu}_{b,i}-j}
}{
1-Q_{f}q^{-\tilde{\mu}_{b,i}+j-1}
}
\prod_{(i,j)\in \tilde{\mu}_{b}}
\frac{
1-q^{\tilde{\mu}_{b,i}-j+1}
}{
1-Q_{f}q^{-\tilde{\mu}_{b,i}+j}
}
,
\nonumber \\ 
\\
&&\Bigl[\partial_{t_{2}}[
\lim_{\epsilon_{2}\to0}
W^{(1)}_{\mu_{b}\tilde{\mu}_{b}}(Q;t,q)W^{(2)}_{\mu_{b}\tilde{\mu}_{b}}(Q;t,q)]\Bigr]
|_{t_{1},~t_{2}~\text{fixed}}
=
\sum_{\mu_{b}}
\frac{
Q_{f}q^{\mu_{b,i}-j+1}
}{
1-Q_{f}q^{\mu_{b,i}-j+1}
}
,
\\
&&\Bigl[ \partial_{t_{f}}[
\lim_{\epsilon_{2}\to0}
W^{(1)}_{\mu_{b}\tilde{\mu}_{b}}(Q;t,q)W^{(2)}_{\mu_{b}\tilde{\mu}_{b}}(Q;t,q)]\Bigr]
|_{t_{1},~t_{2}~\text{fixed}}
=0,
\\
&&\Bigl[\hbar \partial_{\hbar}[
\lim_{\epsilon_{2}\to0}
W^{(1)}_{\mu_{b}\tilde{\mu}_{b}}(Q;t,q)W^{(2)}_{\mu_{b}\tilde{\mu}_{b}}(Q;t,q)]\Bigr]
|_{t_{1},~t_{2}~\text{fixed}}
\nonumber \\
&&=
\hbar
\biggl\{
-
\frac{\mathrm{i}}{2}
\sum_{\mu_{b}}
\frac{
Q_{f}q^{\mu_{b,i}-j}
}{
1-Q_{f}q^{\mu_{b,i}-j}
}
+
\frac{\mathrm{i}}{2}
\sum_{\mu_{b}}
\frac{
Q_{f}q^{\mu_{b,i}-j+1}
}{
1-Q_{f}q^{\mu_{b,i}-j+1}
}
\nonumber \\
&&~~
-
\frac{\mathrm{i}}{2}
\frac{
1
}{
1-Q_{f}q^{-1}
}
\prod_{(i,j)\in \tilde{\mu}_{b},\tilde{\mu}_{b,i}-j\neq0}
\frac{
1-q^{\tilde{\mu}_{b,i}-j}
}{
1-Q_{f}q^{-\tilde{\mu}_{b,i}+j-1}
}
\prod_{(i,j)\in \tilde{\mu}_{b}}
\frac{
1-q^{\tilde{\mu}_{b,i}-j+1}
}{
1-Q_{f}q^{-\tilde{\mu}_{b,i}+j}
}
\biggr\}.
\end{eqnarray}
Then, we can show that the contributions in $F_{\text{M2}}(\bold{t},\hbar)$ which come from the above terms cancel out each other. Thus, we obtain
\begin{eqnarray}
F_{\text{M2}}(\bold{t};\hbar)
=
\frac{\mathrm{i}}{2\pi}\Bigl[
t_{b} \frac{\partial}{\partial t_{b}} \tilde{F}_{\text{NS}}(\mathrm{e}^{-t_{b}};\mathrm{e}^{\mathrm{i}\hbar})
+
\hbar^2 \frac{\partial}{\partial \hbar}\bigl(\tilde{F}_{\text{NS}}(\mathrm{e}^{-t_{b}};\mathrm{e}^{\mathrm{i}\hbar})/h \bigr)
\Bigr]|
+
f(\hbar),
\end{eqnarray}
where we define
\begin{eqnarray}
\tilde{F}_{\text{NS}}(\mathrm{e}^{-t_{b}};\mathrm{e}^{\mathrm{i}\hbar})
=
\lim_{\epsilon_{2}\to 0}
\epsilon_{2}
\mathrm{log}
\biggl[
\sum_{\mu_{b}}
(-Q_{b})^{|\mu_{b}|}
\tilde{Z}_{\mu_{b}}(t,q)\tilde{Z}_{\mu^{t}_{b}}(q,t)q^{\frac{||\mu_{b}||^2}{2}}t^{\frac{||\mu^{t}_{b}||^2}{2}}
\biggr].
\end{eqnarray}
This expression agrees with the non-perturbative parts of the free energy of the resolved conifold except for $f(\hbar)$.
\par
For $\mathcal{Z}_{\emptyset \emptyset}^{(1)}$ and $\mathcal{Z}_{\emptyset \emptyset}^{(2)}$, we can easily show that these contributions become the parts of $f(\hbar)$ by using the following formula,
\begin{eqnarray}
\prod_{i,j=1}^{\infty}(1-Qt^{i-\frac{1}{2}}q^{j-\frac{1}{2}})
=
\mathrm{exp}
\biggl[
-\sum_{m=1}^{\infty}\frac{1}{m}
\frac{Q^m}{(t^{m/2}-t^{-m/2})(q^{m/2}-q^{-m/2})}
\biggr].
\end{eqnarray}

\section{Bubbling Calabi-Yau}
In this section, we show that the partition function of the refined open topological string on the conifold with $N$ toric A-branes agrees with the one of refined closed topological string on the local $\mathcal{B}_{3}$ under some  correspondence. This phenomenon is known as the bubbling Calabi-Yau \cite{Gomis:2006mv}\cite{Taki:2010bj}.
\par
Let us consider the following web diagram.
\begin{figure}[htbp]
\centering
    \includegraphics[width=10cm]{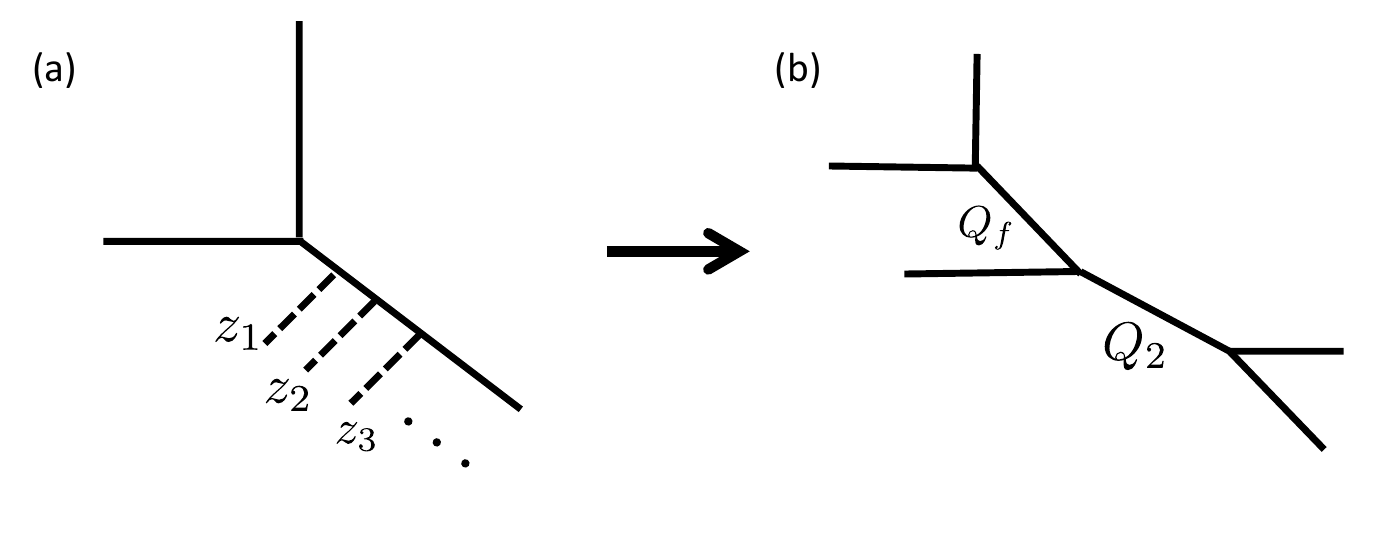}
    \caption{The web diagram.}
    \label{bubbling}
\end{figure}\\
Then, we write the partition function of the refined open topological string by using the refined topological vertex formalism,
\begin{eqnarray}
\mathcal{Z}_{\text{open}}
=
\sum_{\mu}
C_{\emptyset \mu \emptyset}(q,t)\mathrm{Tr}_{{\mu^{t}}}V.
\end{eqnarray}
By using the formula,
\begin{eqnarray}
\mathrm{Tr}_{{\mu}}V=s_{\mu}(x),~V=\mathrm{diag}[z_{1},z_{2},...,z_{N}],
\end{eqnarray}
we obtain
\begin{eqnarray}
\mathcal{Z}_{\text{open}}
=
\prod_{j=1}^{\infty}\prod_{i=1}^{N}
\frac{1
}{
1-z_{i}t^{\frac{1}{2}} q^{j-1}
}.
\label{simple open}
\end{eqnarray}
We next consider the web diagram (b) in Fig.\ref{bubbling}. Then, the partition function of this diagram is 
\begin{eqnarray}
\mathcal{Z}_{\mathrm{closed}}
:=
\sum_{\mu_{f},\mu_{2}}
(-Q_{f})^{|\mu_{f}|}(-Q_{2})^{|\mu_{2}|}
\tilde{f}^{-1}_{\mu_{f}}(t,q)
C_{\mu_{f} \emptyset \emptyset}(q,t)C_{\mu_{2} \mu_{f}^{t} \emptyset}(q,t)
C_{\mu^{t}_{2} \emptyset \emptyset}(t,q).
\end{eqnarray}
After calculation, we obtain
\begin{eqnarray}
\mathcal{Z}_{\mathrm{closed}}
=
\prod_{i,j=1}
\frac{
(1-Q_{2}t^{i-\frac{1}{2}}q^{j-\frac{1}{2}})(1-Q_{2}Q_{f}t^{i-\frac{1}{2}}q^{j-\frac{1}{2}})
}{
(1-Q_{f}t^{i-1}q^{j})
}
\label{simple closed}
\end{eqnarray}
Then, when we set
\begin{eqnarray}
&& Q_{2}=t^N \sqrt{\frac{q}{t}},
 \\
&& Q_{f}qt^{i-\frac{3}{2}}=z_{i}~(i=1,2,...,N).
\end{eqnarray}
The partition function (\ref{simple open}) agrees with (\ref{simple closed}) except for the factors $\prod_{i,j=1}(1-t^{i+N-1}q^{j})$. This difference is due to the difference of normalization between the refined Chern-Simons theory and the refined topological vertex. This is the refinement of the bubbling Calabi-Yau.
\par
The web diagram we consider in section 3 corresponds to the case of $N=1$.

\newpage

\end{document}